\documentclass[1p]{elsarticle}

\usepackage{geometry}
\journal{Physica A}

\usepackage{url}
\usepackage{newcent}
\usepackage[T1]{fontenc}
\usepackage{textcomp}

\usepackage{amssymb}
\usepackage{graphicx}
\usepackage[pdftex,bookmarks,colorlinks,breaklinks]{hyperref}
\hypersetup{linkcolor=red,citecolor=blue,filecolor=dullmagenta,urlcolor=blue}
\usepackage{color}

\usepackage{soul}
\definecolor{indiagreen}{rgb}{0.07, 0.53, 0.03}

\usepackage{mathrsfs}
\usepackage{amsfonts}
\usepackage[fleqn]{amsmath}

\begin{document}


\begin{frontmatter}

\title{Multifractal dimensions and statistical properties of critical %
ensembles characterized by the three classical Wigner--Dyson symmetry classes} %

\author[IFUAP]{M. Carrera-N\'u\~nez}%
\ead{mcarrera@ifuap.buap.mx}%
\author[UNAM]{A.~M. Mart\'inez-Arg\"uello}%
\ead{blitzkriegheinkel@gmail.com}%
\author[IFUAP,USP]{J.~A. M\'endez-Berm\'udez}%
\ead{jmendezb@ifuap.buap.mx}%

\address[IFUAP]{Instituto de F\'isica, Benem\'erita Universidad Aut\'onoma de %
Puebla, Apartado Postal J-48, 72570 Puebla, Mexico}%
\address[UNAM]{Instituto de Ciencias F\'isicas, Universidad Nacional %
Aut\'onoma de M\'exico, Apartado Postal 48-3, 62210, Cuernavaca, Morelos, %
Mexico}%
\address[USP]{Departamento de Matem\'{a}tica Aplicada e Estat\'{i}stica, %
Instituto de Ci\^{e}ncias Matem\'{a}ticas e de Computa\c{c}\~{a}o, %
Universidade de S\~{a}o Paulo - Campus de S\~{a}o Carlos, Caixa Postal 668, %
13560-970 S\~{a}o Carlos, SP, Brazil}%

\begin{abstract}
We introduce a power-law banded random matrix model for the third of the 
three classical Wigner--Dyson ensembles, i.e., the symplectic ensemble. A 
detailed analysis of the statistical properties of its eigenvectors and 
eigenvalues, at criticality, is presented. This ensemble is relevant for 
time-reversal symmetric systems with strong spin--orbit interaction. For the 
sake of completeness, we also review the statistical properties of 
eigenvectors and eigenvalues of the power-law banded random matrix model 
in the presence and absence of time reversal 
invariance, previously considered in the literature. Our results show a good 
agreement with heuristic relations for the eigenstate and eigenenergy 
statistics at criticality, proposed in previous studies. Therefore, we 
provide a full picture of the power-law banded random matrix model 
corresponding to the three classical Wigner--Dyson ensembles.
\end{abstract}


\end{frontmatter}



\section{Introduction}%

The phenomenon of localization of electronic states in disordered mesoscopic 
conductors, first predicted by P. W. Anderson~\cite{Anderson1958}, has 
attracted a lot of theoretical and experimental interest for several 
decades~\cite{Lee1985,Kramer1993}. This phenomenon arises due to quantum 
interference caused by the multiple elastic scattering events of the 
electrons, in their motion along the sample, with randomly distributed 
impurities within the conductor~\cite{Janssen1998}. In the presence of strong 
disorder all electronic states are known to be exponentially localized and 
the conductor behaves as an insulator. Furthermore, as a function of 
parameters like disorder strength, electric or magnetic fields, the sample 
can behave both as an insulator (localized phase) or as a conductor 
(delocalized phase). In addition, the sample can undergo a disorder-induced 
localized--delocalized transition. This transition is usually referred to as 
Anderson or metal--insulator transition (MIT). The understanding of the 
various phenomena that emerge at this transition has been the subject of an 
intense research activity in the last decades~\cite{Lee1985,Janssen1994,Paalanen1991,Sarachick1995,romer2010,romer2011} (for a recent review 
see also the Ref.~\cite{Evers2008} and the references therein). 
A very important 
characteristic feature of this critical point is that not only the electronic 
states, but also the spectra show unusual behavior. For the eigenstates it is 
reflected in multifractal behavior and strong amplitude fluctuations. These 
are usually described by an infinite set of critical 
exponents~\cite{Janssen1998,Janssen1994,Huckestein1995,BHK19,CHR19}.

At the MIT not only the dimensionality, but also the symmetries present in 
the system play an important role. For ordinary disordered samples, the 
random matrix theory (RMT) has proved to be an effective tool in describing 
their statistical properties~\cite{Guhr1998}. Dyson introduced three 
universal symmetry classes: the orthogonal class consisting of systems in 
the presence of time reversal invariance (TRI) and integer spin or TRI, 
half-integer spin, and rotational symmetry; the unitary class describing 
systems with broken time-reversal invariance; and the symplectic class for 
systems in the presence of TRI, half-integer spin, and no rotational 
symmetry. In the Dyson scheme these are labeled by the symmetry indices 
$\beta = 1$, 2, and 4, for the orthogonal, unitary, and symplectic class; 
respectively~\cite{MehtaBook,Dyson1962a,Dyson1962b}.

Up to now, the analysis and theoretical description of the multifractal 
properties of disordered systems at the MIT have been of great 
interest~\cite{Janssen1998,Janssen1994,Evers2008,Bogomolny2011a,Bogomolny2011b,Rushkin2011,Bogomolny2012}. However, because of the 
complexity in obtaining analytical expressions at this critical point, some 
of which are available only perturbatively, many investigations have mainly 
been focused on numerical analysis. In particular a widely used model that 
has attracted a lot of attention is the so-called power-law banded random 
matrix (PBRM) model~\cite{Evers2008,Mirlin1996,Kravtsov1999,Varga2000}. This 
is due to the fact that it captures all the key features of the Anderson 
critical point and is also well suited for numerical calculations.

More recently~\cite{Antonio2012,Antonio2014}, this model has been used to 
verify the validity of existing heuristic relations, established between the 
multifractal properties of eigenstates and their spectra at criticality. 
These relations have been proved to correctly describe the multifractal 
behavior of critical states for the PBRM model in the presence ($\beta=1$)
~\cite{Antonio2012} and absence ($\beta=2$)~\cite{Antonio2014} of time 
reversal invariance, in a relatively wide range of the model parameters. 
Furthermore, it also accounts for the multifractal properties of other 
models showing critical behavior~\cite{Antonio2012,Antonio2014}. However, 
the PBRM model corresponding to the third of the three classical 
Wigner--Dyson ensembles, i.e.,~the symplectic case, has been left out. This 
model is of particular interest since it has been shown, using 
two-dimensional tight-binding models, the appearance of a 
localized--delocalized transition in systems that belong to the symplectic 
symmetry~\cite{Evers2008,Evangelou1987,Evangelou1995,Fastenrath1991,Fastenrath1992,Chalker1993,Schweitzer1997,ohtsuki1995,Markos2006}.

Our purpose in this paper is to introduce the PBRM model for the symplectic 
class, i.e.,~for time reversal symmetric systems in the presence of 
spin--orbit interactions. For the sake of completeness, we also review the 
corresponding multifractal and statistical properties of systems in the 
presence and absence of time reversal invariance so that we can provide a 
full picture of the PBRM model for the three classical Wigner--Dyson ensembles. 
We show that the heuristic relations, proposed earlier~\cite{Antonio2012,Antonio2014}, for multifractal properties are also valid for the
symplectic 
case, in certain range of the model parameters.

The paper is organized as follows. In the next Section the PBRM model in the %
presence of the three Wigner--Dyson symmetries is described. There, we review %
the PBRM model for $\beta=1$ and 2, and introduce the model for the %
symplectic class, i.e., the one with $\beta=4$. In %
Section~\ref{sec:relations} and \ref{subsec:eigenstates} we present the %
heuristic relations for the eigenstates statistics of the PBRM at criticality. %
The numerical results for the eigenstates statistics are presented in %
Section~\ref{subsec:vec}. The spectral analysis is the subject of %
Section~\ref{sec:SpectralAnalysis} and in Section~\ref{subsec:ev} we discuss %
the results. Finally, we present our conclusions in %
Section~\ref{sec:conclusions}.%


\section{The PBRM model for the Wigner--Dyson ensembles}%
\label{sec:models}%
We start our study by defining the power-law banded random matrix (PBRM) 
model in the presence of the three symmetry classes characterized by the 
classical Wigner--Dyson ensembles. The PBRM model describes one-dimensional 
($1d$) samples with random long-range hoppings. This model is represented by 
$N\times N$ real symmetric ($\beta=1$), complex Hermitian ($\beta=2$), or 
$2N\times 2N$ quaternion-real Hermitian ($\beta=4$) matrices whose elements 
are statistically independent random variables drawn from a normal 
distribution with zero mean and variance given by
\begin{align}
\label{eq:ModelB}
\langle |H_{mm}|^{2} \rangle & =  \beta^{-1} \quad \mathrm{and} \nonumber \\
\langle |H_{mn}|^2\rangle & =  \frac{1}{2(1 + \delta_{\beta, 4})}\frac{1}{ 1+
\left[ \sin\left(\frac{\pi|m-n|}{N}\right)\Big/
\left(\frac{\pi b}{N}\right) \right]^{2\mu}} ,
\end{align}
where $\mu$ and $b$ are the model parameters. The model of~\eqref{eq:ModelB} is in its periodic version; i.e.,~the $1d$ sample 
is in a ring geometry. 
In this paper, the Hamiltonian matrix 
$H$ of the PBRM model for the symplectic case ($\beta=4$) is a Hermitian 
self-dual matrix, i.e.,~$H=H^{\dagger}$ and $H=H^{\mathrm{R}}$ 
with $H^{\dagger}$ and $H^{\mathrm{R}}$ the adjoint and 
dual of $H$, respectively. Its elements $H_{mn}$ are given 
by~\cite{MehtaBook}
\begin{equation}
H_{mn} = \sum_{\alpha=0}^{3} q_{mn}^{(\alpha)} \tau_{\alpha},
\end{equation}
which are written in terms of the $2\times 2$ unit matrix, %
$\tau_{0}=\mathbf{1}_{2}$, and the quaternions %
\begin{equation}
\label{eq:quaternions}
\tau_{1} = \left(
\begin{array}{cc}
i & 0 \\
0 & -i 
\end{array}
\right), \quad
\tau_{2} = \left(
\begin{array}{cc}
0 & 1 \\
-1 & 0 
\end{array}
\right), \quad \mathrm{and} \quad
\tau_{3} = \left(
\begin{array}{cc}
0 & i \\
i & 0 
\end{array}
\right),
\end{equation}
and $q^{(\alpha)}_{mn}$ are real numbers. This model 
is introduced in the same line as the one originally proposed 
by Mirlin \textit{et al.} for $\beta=1$~\cite{Mirlin1996}. That is, a random 
matrix ensemble of the symplectic class with off-diagonal matrix elements 
decaying away from the diagonal in a power-law fashion with zero mean and 
variance given by~\eqref{eq:ModelB}. It is worth mentioning that 
such a model also preserves the quaternion structure of the Hamiltonian 
where each eigenvalue is two-fold degenerate due to Kramers degeneracy.

The PBRM model, as shown in~\eqref{eq:ModelB}, depends on the two 
control parameters $\mu$ and $b$. The power-law decay $\mu$ induces a 
disorder-driven  MIT~\cite{Evers2008,Mirlin1996,Varga2000,Mirlin2000,Kravtsov1997,Kravtsov2000,Cuevas2001,Varga2002}: when $\mu > 1$
($\mu < 1$) 
the PBRM model is in the insulating (metallic) phase, so its eigenstates are 
localized (delocalized); here the MIT occurs at $\mu = 1$. Indeed, it has 
been shown~\cite{Evers2008,Antonio2012,Antonio2014,Antonio2005,Antonio2006a,Antonio2006b,Antonio2009,Alcazar2009} that the PBRM model with
$\beta=1$ and 
$\beta=2$ exhibits a broad spectrum of critical properties at the MIT. 
Regardless of the value of $\mu$, the effective bandwidth $b$ remains as a 
free parameter able to tune the localization properties of the PBRM model. 
At the MIT, the eigenstates of the PBRM model may transit from strong 
multifractals to weak multifractals~\cite{Evers2008,Mirlin2000} by moving the 
bandwidth from $b \ll 1$ to $b \gg 1$, respectively, with $b \in (0, \infty)$.

In this paper, we show that the critical point for the symplectic case %
also occurs at $\mu= 1$ (see~\ref{sec:appendixA}), and focus on the PBRM %
model in the presence of the three symmetry classes labeled by %
$\beta=1, 2,$ and 4, at the MIT point $\mu = 1$. %


\section{Heuristic relations}%
\label{sec:relations}%

Recently, heuristic relations between the multifractal properties of %
eigenstates and their spectra at criticality have been proposed and verified %
for the PBRM model with $\beta=1$ and 2~\cite{Antonio2012,Antonio2014}. Here, %
we review these relations and show that they also account for the multifractal %
properties and the spectra of the PBRM model in the presence of the %
symplectic symmetry ($\beta=4$), proposed in equation~(\ref{eq:ModelB}).%


\subsection{Multifractality of electronic states}%
\label{subsec:eigenstates}%

It is widely known that the spatial fluctuations of electronic states in %
disordered conductors at the Anderson transition show multifractal %
behavior~\cite{Janssen1998,Janssen1994,Evers2008,Hashimoto2008,Faez2009,%
Richardella2010}. These fluctuations can be described by a set of %
multifractal dimensions $D_{q}$ defined by the scaling of the inverse mean %
eigenfunction participation numbers with the system size $N$:
\begin{equation}
\label{eq:invpn}
\left\langle \sum_{i=1}^N|\Psi_{i}|^{2q}\right\rangle \sim
N^{-(q-1)D_{q}},
\end{equation}
where $q$ is a parameter and $\langle\cdots\rangle$ stands for the average %
over some eigenstates within an eigenvalue window and over the ensemble. %
Notice that, the so-called information dimension, $D_{1}$, can not be computed %
from equation~(\ref{eq:invpn}). However, in the limit %
$q\rightarrow1$, $D_{1}$ is related to the information entropy of the %
eigenstates as
\begin{equation}
\label{eq:invpn1}
\left\langle -\sum_{i=1}^N|\Psi_{i}|^{2}\ln|\Psi_{i}|^{2}\right\rangle \sim
D_{1}\ln N.
\end{equation}

For the PBRM model of equation~(\ref{eq:ModelB}), in both the presence %
($\beta=1$)~\cite{Antonio2012} and absence ($\beta=2$)~\cite{Antonio2014} of time %
reversal invariance, a heuristic relation for the multifractal dimension, given by%
\begin{equation}
\label{eq:heudq}
D_{q}\approx[1+(\alpha_{q}b)^{-1}]^{-1},
\end{equation}
has been shown to be valid in a wide range of the parameters $q$ and $b$. %
Here, $\alpha_{q}$ is a fitting constant. %

In addition, an often employed quantity to %
characterize the spectral fluctuations is the level compressibility, $\chi$, %
obtained from the limiting behavior of the spectral number variance as %
$\sum^{(2)}(E) = \langle n(E)^2\rangle-\langle n(E)\rangle^2 \sim \chi\langle %
n(E)\rangle$ %
where $n(E)$ is the number of eigenstates in an interval of length $E$ and
$\langle n(E)\rangle\gg1$. %
In the metallic regime, where the states are extended, $\chi=0$; %
while in a strongly disordered conductor the levels are uncorrelated and lead %
to $\chi=1$. Furthermore, at criticality an intermediate %
statistics exists, $0< \chi <1$, where the spectral and eigenstate statistics %
are supposed to be coupled~\cite{Chalker1996,Klesse1997}. Analytical %
expressions, for $\beta=1$ and 2, that describe qualitatively well the %
level compressibility, $\chi$, in the small- and large-$b$ limits are %
given by~\cite{Evers2008,Kravtsov2006,Kravtsov2011}%
\begin{equation}
\label{eq:chi}
\chi=
\left\{
\begin{array}{lr}
1 - 4b,~~~~~~~~~~~~\;~~~~~~~~~~~~~~~~\beta=1, & b\ll1 , \\
1-\pi\sqrt{2}b+\frac{4}{3}(2-\sqrt{3})\pi^2 b^2, \, \, \, \beta=2 , & b\ll1 , \\
\frac{1}{2\beta\pi b}, &  b\gg1 ,
\end{array} \right.
\end{equation}
which can be written in terms of the multifractal dimensions, $D_{q}$, %
as~\cite{Antonio2014}%
\begin{equation}
\label{eq:chiapprox}
\chi\approx\frac{1-D_q}{1+(q-1)D_{q}}.
\end{equation}
Moreover, the multifractal dimensions, $D_{q}$, can also be related to the %
information dimension, $D_{1}$, as~\cite{Antonio2014}:%
\begin{equation}
\label{eq:DqofD1}
D_{q}\approx\frac{D_{1}}{q+(1-q)D_{1}}.
\end{equation}

It is worth mentioning that equations~(\ref{eq:heudq}), (\ref{eq:chiapprox}), %
and (\ref{eq:DqofD1}) are valid only for $q>1/2$. However, for the case in %
which $q<1/2$, $D_{q}$ and $D_{1}$ are related as~\cite{Antonio2014}%
\begin{equation}
\label{eq:Dqless}
D_{q}\approx\frac{1-2q}{1-q}+\frac{q}{1-q} \frac{D_1}{1+q(D_1-1)} .
\end{equation}
Thus, from equations~(\ref{eq:DqofD1}) and (\ref{eq:Dqless}) it is possible %
to explore the whole range of $q$.%


\subsection{Numerical results for multifractality}
\label{subsec:vec}

We now verify the heuristic relations presented so far. For the numerical %
analysis of the eigenstates we use system sizes of $N=2^{n}$, with $8\leq %
n\leq 13$. The averages are taken over $12.5\%$ of the eigenvectors within an %
eigenvalue window around the band center of the spectrum, and over ensembles %
of sizes $M=2^{16-n}$. This guarantees that the statistics is fixed, i.e., %
the product $N\times M=2^{16}$ for any system and ensemble size. %
The reported %
error bars are the reduced rms error of the fittings between the numerical 
data and the corresponding analytical expression. %


\begin{figure*}
\centering
\includegraphics[width=0.67\textwidth]{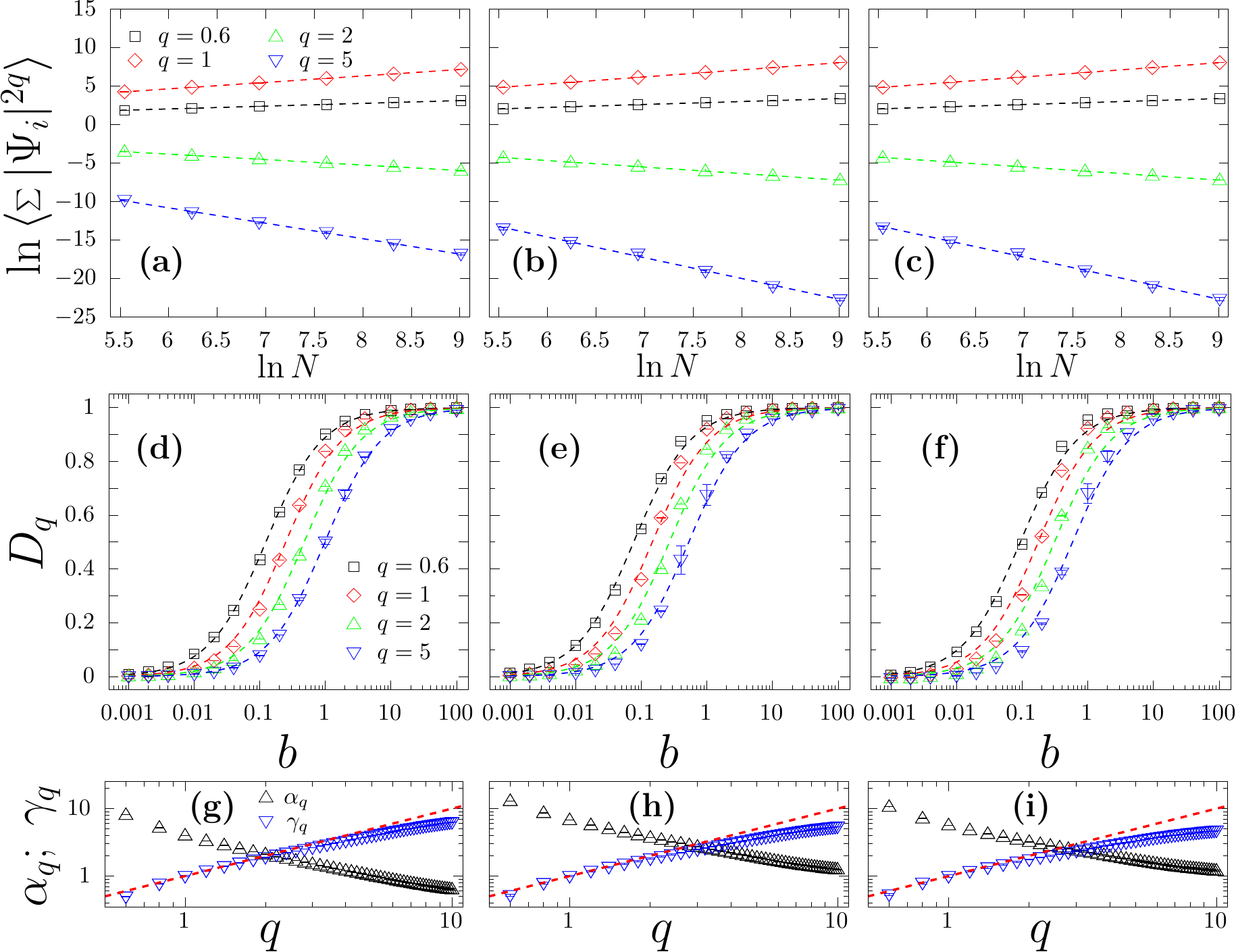}
\caption{Eigenstate statistics of the PBRM model. Left column %
PBRM model with $\beta=1$ [panels (a), (d) and (g)], middle column with %
$\beta=2$ [panels (b), (e) and (h)], and right column with $\beta=4$ %
[panels (c), (f) and (i)]. See text for a detailed description.}%
\label{fig:eigenst1}%
\end{figure*}


In Figure~\ref{fig:eigenst1}, panels (a)-(c), we show the logarithm of the %
mean generalized inverse participation numbers of equation~(\ref{eq:invpn}) %
as a function of the logarithm of the system size, for several values of $q$. %
In all these cases the parameter $b$ is fixed to one. We %
observe that equation~(\ref{eq:invpn}) is in complete agreement with the %
numerical results of the PBRM model in the presence of the three symmetry %
classes, shown in panels (a), (b), and (c), for $\beta=1$, 2, and 4, %
respectively. Also, from these results we extract the multifractal %
dimensions, $D_{q}$, by performing a linear fitting to the numerical data, %
as shown in dashed lines of the same figures. For the special case of %
$q=1$ we used equation~(\ref{eq:invpn1}).%

Now in Figure~\ref{fig:eigenst1}, panels (d)-(f), we show $D_{q}$ as a %
function of $b$ for the same values of $q$ considered previously. There, the %
dashed lines correspond to fittings to the heuristic relation given by %
equation~(\ref{eq:heudq}). We observe that equation~(\ref{eq:heudq}) is in %
agreement with the PBRM model with $\beta=1$ and 2 [see panels (d) and (e)]. %
For the $\beta=4$ case, we observe some deviation from equation%
~(\ref{eq:heudq}) specially for values of $b$ between $0.04<b<2$ where %
$D_{q}$ grows faster than expected. However, in the limiting cases $b\ll1$ %
(insulator phase) and $b\gg1$ (metallic phase) the behavior of $D_{q}$ is %
well described by equation~(\ref{eq:heudq}). These results have already been %
reported for the cases of $\beta=1$~\cite{Antonio2012} and $\beta=2$%
~\cite{Antonio2014}.%

The coefficients $\alpha_{q}$ of equation~(\ref{eq:heudq}) as a function %
of $q$ are displayed in panels (g)-(i) of Figure~\ref{fig:eigenst1} in %
black-triangles. The cases $\beta=1$ (g) and $2$ (h) have been reported in %
Refs.~\cite{Antonio2012} and \cite{Antonio2014}, respectively. The $\beta=4$ %
case is shown in panel~(i). There, we observe an almost linear decay of 
$\alpha_{q}$ as $q$ %
increases with values that lie between the cases $\beta=1$ and $\beta=2$. %
Also, in panels (g)-(i) of the same figure, we show the quantity %
$\gamma_{q}=\alpha_{1}/\alpha_{q}$ as a function of $q$ in %
blue-inverted-triangles. The red-dashed lines correspond to $\gamma_{q}=q$. %
This is an interesting result since in the region $0<q<2.5$ we can calculate %
these coefficients as $\alpha_{q}\approx\alpha_{1}/q$ with high accuracy for %
the PBRM model in the presence of the three symmetry classes $\beta=1$, 2, and %
4, and then obtain very simplified recursive relations for several %
interesting quantities such as the multifractal dimension $D_{q}$ as %
given in equation~(\ref{eq:heudq}).%

\begin{figure*}
\centering
\includegraphics[width=0.70\textwidth]{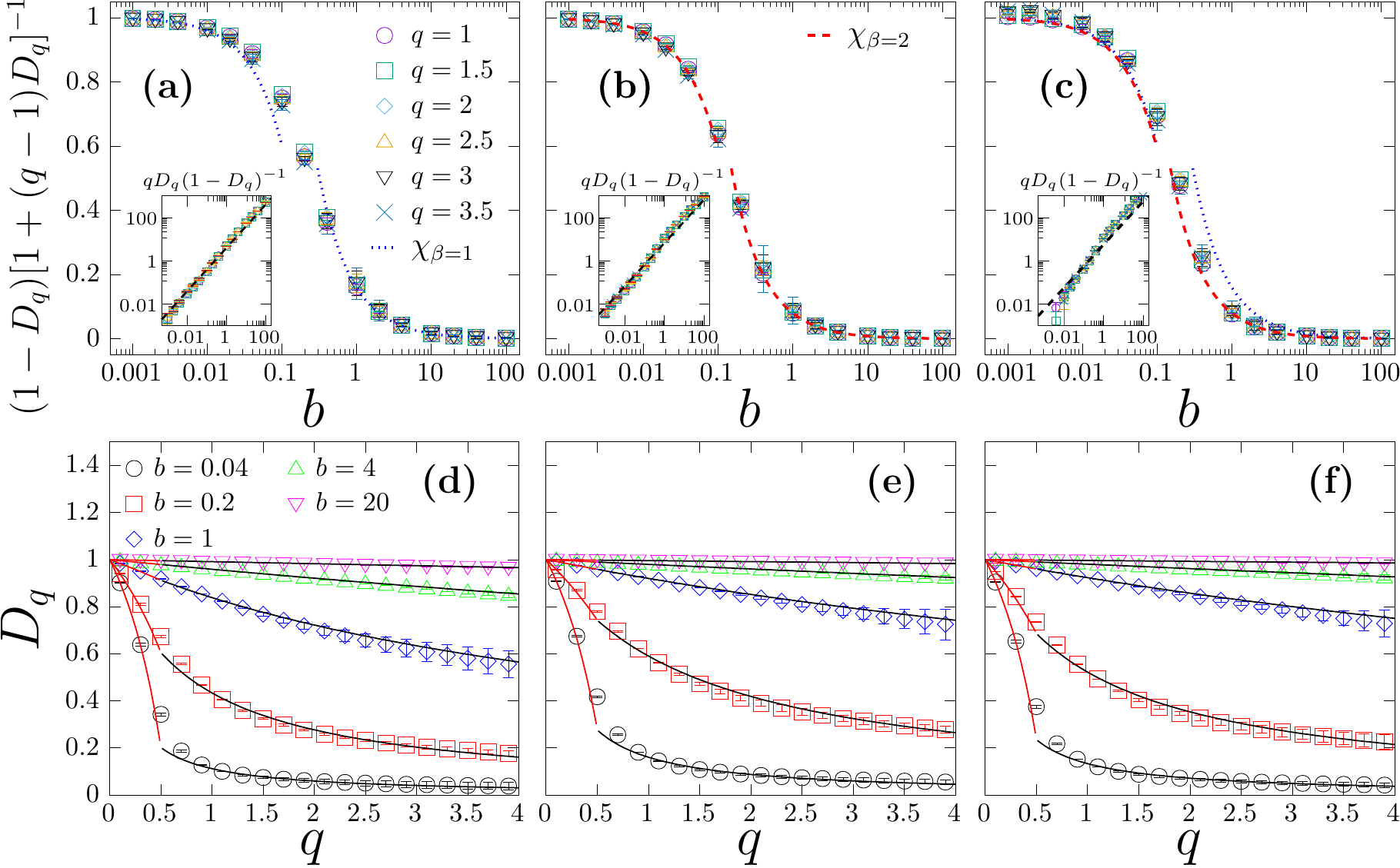}
\caption{Eigenstate multifractal dimensions of the PBRM model. %
Left column, panels (a) and (d), corresponds to the PBRM model with $\beta=1$. %
Middle column, panels (b) and (e), corresponds to the PBRM model with %
$\beta=2$. Right column, panels (c) and (f), corresponds to the PBRM model %
with $\beta=4$. See text for a detailed description.}%
\label{fig:eigenst2}
\end{figure*}

In Figure~\ref{fig:eigenst2}, panels (a)-(c), we %
verify the validity of the heuristic relation of equation~(%
\ref{eq:chiapprox}), compared to the level compressibility $\chi$ %
given by the analytical expression of equation~(\ref{eq:chi}), for %
several values of $q$ as indicated in the figure. %
Equation~(\ref{eq:chiapprox}) has already %
been validated for $\beta=1$ in Ref.~\cite{Antonio2012} and for $\beta=2$ in %
Ref.~\cite{Antonio2014}. For the sake of completeness we reproduce these %
results, shown in Figure~\ref{fig:eigenst2} in panels (a) and (b), for $\beta =1$ %
and 2, respectively; the results for $\beta=4$ are given in %
Figure~\ref{fig:eigenst2} (c). %
The blue-dotted line in panel (a) and the red-dashed one in panel (b) %
correspond to the analytical expression of equation~(\ref{eq:chi}), where for %
the sake of clarity we have added the subindex $\beta=1$ and $\beta=2$ %
according to the case considered. For comparison purposes, we have included %
both analytical expressions for $\chi$ for the cases $\beta=1$ (blue-dotted %
line) and $\beta=2$ (red-dashed line), together with the numerical result for %
the symplectic case (dots) in panel (c). There, we can see that when $b<0.1$ %
both curves, for $\beta=1$ and 2, show the same behavior, which is different %
in the symplectic case (dots). Furthermore, when $b>1$ the symplectic case %
behaves quite similar to $\beta=2$, and when $b>10$ the three models show the %
same behavior, which is expected since they have reached the metallic phase. %
An interesting region is in the interval between $0.04<b<0.4$ where the three %
models show a different behavior. This matches with the region where $D_{q}$ %
changes fast [see Figure~\ref{fig:eigenst1}, panels (d)-(f)]. In insets of %
panels (a)-(c) of Figure~\ref{fig:eigenst2} we show $qD_{q}(1-D_{q})^{-1}$ %
(dots) as a function of $b$. The red-dashed lines correspond to $\alpha_{1}b$. %
From those figures we can see that $qD_{q}(1-D_{q})^{-1}\approx\alpha_{1}b$ %
and therefore it does not dependent on $q$. This explains why all numerical %
data shown in panels (a)-(c) of Figure~\ref{fig:eigenst2} fall almost on the %
same point for a given $b$.%


Finally, we present the results for the multifractal dimension, $D_{q}$, for %
$q<1/2$ and $q>1/2$ for different values of the parameter $b$. These are %
shown in panels (d)-(f) of Figure~\ref{fig:eigenst2} for the three symmetry %
classes $\beta=1$, 2, and 4, respectively. The red lines correspond to %
expression~(\ref{eq:Dqless}) while the black ones correspond to %
relation~(\ref{eq:DqofD1}). Again, the results for $D_{1}$ are obtained from %
equation~(\ref{eq:invpn1}). As can be seen the results show a good agreement %
between the numerical simulations and the theoretical expressions for the %
PBRM model, confirming the validity of those heuristic relations for the %
three symmetry classes analyzed here.%


\section{Spectral statistics at criticality}%
\label{sec:SpectralAnalysis}%

In the previous section we have analyzed the eigenvector statistics of the %
PBRM model at the critical point. Such study was done through the multifractal %
dimensions and the level compressibility. Another effective tool to distinguish %
between localized and extended states is given by the nearest level spacing %
distribution. This distribution also accounts for the symmetry class present %
in the system. Since the PBRM model allows a transition from localized to %
extended states by tuning the parameter $b$, in this section we shall analyze %
the spectral statistics of the PBRM model at criticality as a function of $b$.%


\subsection{The nearest-neighbor level spacing distribution}%

The spectrum of a disordered system in the localized regime (insulator phase) %
is uncorrelated and follows a Poisson nearest-neighbor level spacing %
distribution~\cite{Berry1977}%
\begin{equation}
\label{eq:poissonlaw}
P_{\mathrm{P}}(s)=\mathrm{exp}(-s),
\end{equation}
where $s=|e_{i+1}-e_{i}|/\Delta$, with $e_{i}$ the %
eigenenergies and $\Delta$ the mean level spacing. In contrast, %
to a good approximation, the %
spectrum of a disordered system in the delocalized regime %
(metallic phase) follows a nearest-neighbor level %
spacing distribution of one of the three Wigner--Dyson surmises %
~\cite{Bohigas1984}%
\begin{equation}
\label{eq:wignerdyson}
P_{\rm WD}(s)=
\left\{
\begin{array}{lr}
\frac{\pi}{2}s\exp\left(-\frac{\pi}{4}s^2 \right), & \beta = 1 , \\
\\
\frac{32}{\pi^2}s^2\exp\left(-\frac{4}{\pi}s^2 \right), & \beta = 2 , \\
\\
\frac{2^{18}}{3^{6}\pi^3}s^4\exp\left(-\frac{64}{9\pi}s^2 \right), & \beta = 4,
\end{array} \right.
\end{equation}
depending of the symmetry present in the system %
labeled by the Dyson symmetry index $\beta$. Now, the PBRM model allows us to %
study the signatures of the metal-insulator phase transition in its spectra. %
However, the spectrum of the PBRM model follows a more general distribution %
$P_{\mathrm{c}}(s)$ which is neither Poissonian nor Wigner--Dyson, but in the %
limiting cases it agrees with equation~(\ref{eq:poissonlaw}) when %
$b\rightarrow 0$ and with equation~(\ref{eq:wignerdyson}) when %
$b\rightarrow \infty$. Since the spectral statistics shows a different %
behavior for small and large differences in energy, $s$, we divide our %
spectral analysis into two parts: the first one for $s\gg1$ and the other one %
for $s\ll1$.%

In the case of large energy differences, $s\gg1$, it has been derived %
analytically~\cite{Aronov1994,Aronov1995,Kravtsov1995}, and numerically %
verified~\cite{Cuevas2004} %
for $\beta=1$, that $P_{\rm c}(s)$ has the following asymptotic form%
\begin{equation}
\label{eq:Pcs1}
P_{\rm c}(s)\sim\exp\left(-As^\alpha \right) ,\quad s\gg1,
\end{equation}
where $A$ is a coefficient that depends only on the %
dimensionality of the system, and %
$\alpha$ is the characteristic exponent. For $\beta=1$, $\alpha$ ranges in the %
interval $1<\alpha<2$ and it has been conjectured that can be fitted according to~\cite{Cuevas2004}%
\begin{equation}
\label{eq:alphafit}
\alpha =
\left\{
\begin{array}{lr}
2-a/b , & b\gg1 , \\
1+cb , & b\ll1 ,
\end{array} \right.
\end{equation}
with $a$ and $c$ constants. In order to diminish the magnitude of the %
relative fluctuations we do not consider the nearest-neighbor level spacing %
distribution directly, but the cumulative level spacing distribution function%
\begin{equation}
\label{eq:cummulative}
I(s)=\int_{s}^{\infty}P(s'){\rm d}s'.
\end{equation}
Meanwhile, in the case of small energy differences, $s\ll1$, the nearest%
-neighbor level spacing distribution $P_{\rm c}(s)$ behaves %
as~\cite{MehtaBook}%
\begin{equation}
\label{eq:pcs2}
P_{\rm c}(s)\sim C\, s^{\beta} ,\quad s\ll1,
\end{equation}
where $C$ is a constant to be determined and $\beta$ is the Dyson symmetry %
index.%

Based on a random-matrix model with log-squared potentials of the form %
$(1/a)(\ln x)^2$, being $a$ a parameter of the model, %
Nishigaki~\cite{Nishigaki1998,Nishigaki1999} was %
able to provide analytical estimates for the full range of the level spacing %
distribution function $P_{\beta}(s)$ at the critical point:%
\begin{align}
P_{1}(s)&=\left[e^{-(1/2)\int_{0}^{s}ds[R(s)+\sqrt{R'(s)}]}\right]''\; ,
\label{eq:P1} \\
P_{2}(s)&=\left[ e^{-\int_{0}^{s}ds R(s) }\right]''\; , \label{eq:P2} \\
P_{4}(s)&=\left[ e^{-(1/2)\int_{0}^{2s}ds R(s)}
\cosh\left(\frac{1}{2}\int_{0}^{2s}ds\sqrt{R'(s)} \right) \label{eq:P4}
\right]''\; ,
\end{align}
in which the parameter $a$ is such that $\mathrm{exp}(-\pi^{2}/a) \ll 1$, %
the primes $(')$ indicate derivation with respect to $s$ and $R(s)$ %
obeys the following differential equation%
\begin{equation} 
\left[a\cosh as R'(s)+\frac{\sinh as}{2}R''(s)\right]^2 + 
[\pi
\sinh as  R'(s)]^2 
=
R'(s)\left( [aR(s)]^2 \right. 
\left. +a\sinh 2as R(s) R'(s) +[\sinh as R'(s)]^2 \right).
\label{eq:diffeq1}
\end{equation}
The boundary condition of the above equation can be found from %
$R(s)=1+s+\mathcal{O}(s^2)$. Furthermore, for $s \ll 1/a$ a perturbative %
solution to equation~(\ref{eq:diffeq1}) yields (see Refs.~\cite{Nishigaki1998,%
Nishigaki1999} for details)%
\begin{align}
R(s)&=1+s+s^2+\left( 1-\frac{\pi^2+a^2}{9}\right)
s^{3}+\left( 1-\frac{5(\pi^2+a^2)}{36}\right)s^{4}
+ \left( 1-\frac{(\pi^2+a^2)(75-4\pi^{2}-6a^{2})}{450}\right)s^{5}+\ldots,
\label{eq:perR}\\[0.5cm]
P_{1}(s)&=\frac{4\pi^{2}+a^{2}}{24}s
-\frac{(4\pi^{2}+a^{2})(12\pi^{2}+7a^{2})}{2880}s^{3} 
+\frac{(\pi^{2}+a^{2})(4\pi^{2}+a^{2})}{1080}s^{4} 
+\frac{(4\pi^{2}+a^{2})(48\pi^{4}+
                          72\pi^{2}a^{2}+31a^{4})}{322560}s^{5}\nonumber \\
&\;\;~-\frac{(\pi^{2}+a^{2})(4\pi^{2}+a^{2})(12\pi^{2}+13a^{2})}{226800}s^6
                          +\ldots,\label{eq:perP1}\\[0.5cm]
P_{2}(s)&=\frac{\pi^{2}+a^{2}}{3}s^{2}
-\frac{(\pi^{2}+a^{2})(2\pi^{2}+3a^{2})}{45}s^{4}
+\frac{(\pi^{2}+a^{2})(\pi^{2}+2a^{2})(3\pi^{2}+5a^{2})}{945}s^{6}+\ldots,
\label{eq:perP2}\\[0.5cm]
P_{4}(s)&=\frac{16(\pi^{2}+a^{2})(\pi^{2}+4a^{2})}{135}s^{4}
-\frac{128(\pi^{2}+a^{2})(\pi^{2}+4a^{2})(3\pi^{2}+13a^{2})}{14175}s^{6}+
\ldots.\label{eq:perP4}
\end{align}
These expressions have been successfully validated with the tight-binding %
Anderson Hamiltonian with $\beta=1$, 2, and 4~\cite{Nishigaki1999}.%


\subsection{The ratio of consecutive level spacings distribution}%
\label{subsec:ratio}%

The main disadvantage in calculating the nearest-neighbor level spacing %
distribution, equation~(\ref{eq:wignerdyson}), is the need to perform the %
unfolding procedure, which in some cases is not possible~\cite{Guhr1998}, for %
example in many-body problems. To overcome these difficulties new quantities %
have been proposed. For instance, for an ordered spectrum $\{e_{n} \}$ from a %
random matrix the nearest-neighbor spacing is $s_{n}=(e_{n+1}-e_{n})$ and the %
ratio of consecutive level spacings is defined by~\cite{Oganesyan2007}%
\begin{equation}
\label{eq:eratioconsec}
\tilde{r}_{n}=\frac{\min(s_{n},s_{n-1})}{\max(s_{n},s_{n-1})} = \min\left(r_{n},\frac{1}{r_{n}} \right),
\end{equation}
where
\begin{equation}
\label{eq:eratioconsec2}
r_{n}=\frac{s_{n}}{s_{n-1}}.
\end{equation}
These quantities, $r_{n}$ and $\tilde{r}_{n}$, have the advantage of requiring %
no unfolding, and therefore can be compared directly with experimental data or %
with a true spectrum. It is %
known that $r$ has a closed probability density function (PDF) for the three %
classical Wigner--Dyson ensembles~\cite{Atas2013,Atas2013b}%
\begin{equation}
\label{eq:PWr}
P_{\mathrm{WD}}(r)=\frac{1}{Z_{\beta}}
\frac{(r-r^2)^{\beta}}{(1+r+r^{2})^{1+3\beta/2}},
\end{equation}
where $Z_{\beta}$ is a normalization constant given by%
\begin{equation}
\label{eq:zbeta}
Z_{\beta} = \frac{2\pi \Gamma(1 + \beta)}{3^{3(1 + \beta)/2}\, [\Gamma(1 + \beta/2)]^2} ,
\end{equation}
being $\Gamma$ the usual Gamma function and $\beta$ the Dyson symmetry index. %
A closed form for the PDF of $\tilde{r}$ for the Wigner--Dyson ensembles is not %
available up to now. However, for the integrable case (Poisson) these are~%
\cite{Atas2013}%
\begin{equation}
\label{eq:prpoisson}
P_{\rm P}(r)=\frac{1}{(1+r)^{2}},
\end{equation}
and%
\begin{equation}
\label{eq:ptrpoisson}
P_{\rm P}(\tilde{r})=\frac{2}{(1+\tilde{r})^{2}}.
\end{equation}

In what follows, we present our numerical results for the spectral statistics %
for the PBRM model in the presence of the three symmetry classes, $\beta=1$, 2, %
and~4.%


\subsection{Numerical results for the spectral statistics}%
\label{subsec:ev}%

In this section, we present the numerical results concerned to the different %
aspects of the spectrum of the PBRM model in the presence of the three %
symmetry classes $\beta=1$, 2, and 4, at the metal-insulator transition as %
defined in equation~(\ref{eq:ModelB}). For our numerical simulations we %
consider system sizes of $N=2^{n}$, with $n=6$ (red-diamonds), $7$ (green-%
triangles), and $8$ (blue-circles). The ensemble sizes $M$ are chosen such %
that the product $N\times M$ remains fixed to $5.12\times10^{7}$. Also, we %
consider eigenvalue windows within the $12.5\%$ around the center of the %
spectrum. As before, the error bars in the different panels %
are the reduced rms error of the fittings between the heuristic predictions %
and the numerical data. %


\begin{figure*}[t]
\begin{center}
\includegraphics[width=0.70\textwidth]{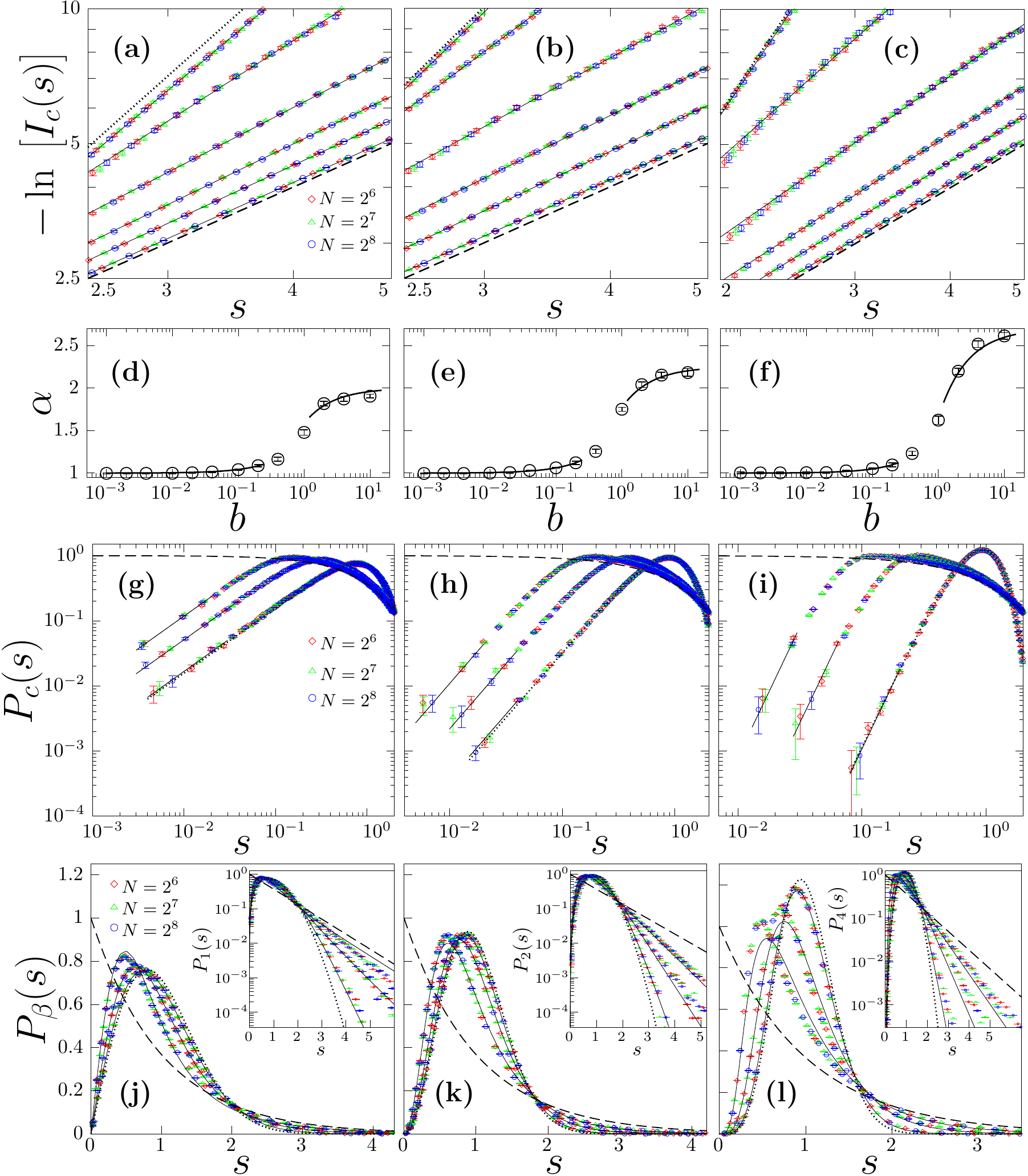}
\caption{%
Spectral statistics of the PBRM model for $\beta=1$ in %
left column, panels (a), (d), (g), and (j); for $\beta=2$ in middle column, %
panels (b), (e), (h), and (k); and for $\beta =4$ in right column, panels (c), %
(f), (i), and (l). %
The dashed lines correspond to the Poisson case, equation~(\ref{eq:poissonlaw}) %
while the dotted ones represent the corresponding Wigner--Dyson surmises of %
equation~(\ref{eq:wignerdyson}), in all panels. In panels (a)-(c) the %
values of the parameter $b$ are: $b=0.02,0.1,0.2,0.4,1$ and $4$ (from bottom %
to top). In panels (g)-(i) we have $b=0.04,0.1$ and $4$ (from left to right). %
In main panels (j)-(l) the values of $b$ are: $b=0.2,0.4$ and $1$ (from left to right).%
}%
\label{fig:spectral1}
\end{center}
\end{figure*}


In Figure~\ref{fig:spectral1}, panels (a)-(c), are plotted the minus logarithm %
of the integrated probability, equation~(\ref{eq:cummulative}), for the PBRM %
model with $\beta=1$, 2, and 4, respectively. In those panels we show the %
results for the following values of parameter $b$: $0.02,\;0.1,\;0.2,\;0.4,\;%
1$, and $4$, from bottom to top. The dashed lines correspond to the Poisson %
case, equation~(\ref{eq:poissonlaw}), while the dotted lines are %
the corresponding Wigner--Dyson surmises, equation~(\ref{eq:wignerdyson}). The %
solid lines are fittings to equation~(\ref{eq:Pcs1}). The result for $\beta=1$ %
is in agreement with that reported in~\cite{Cuevas2004}, while the results for %
the cases $\beta=2$ and $\beta=4$ have not been reported so far. Here, we %
verified that for both cases, $\beta=2$ and 4, the %
coefficient $A\approx 1$, and %
the fittings are done in the range of $s$ shown in each one of those panels, %
i.e., for instance, if $b<0.2$ we have $2.5<s<5$ for $\beta=2$, and $2<s<5$ %
for $\beta=4$, and so on. With these results we confirm the asymptotic %
behavior of $P_{\rm c}(s)$ predicted in equation~(\ref{eq:Pcs1}).%

The characteristic exponent, $\alpha$, as a function of $b$ %
[see equation~(\ref{eq:alphafit})] is plotted in panels (d)-(f) of Figure~\ref{fig:spectral1}. %
The results for the $\beta=1$ case, panel (d), are in agreement with those %
reported in~\cite{Cuevas2004}. In the same panel, the line corresponds to the %
fitting of equation~(\ref{eq:alphafit}) to the numerical data. Although equation~(\ref{eq:alphafit}) %
was conjectured to be valid only for the $\beta=1$ case, it gives an accurate %
description of the numerical data for the $\beta=2$ and $\beta=4$ cases %
for $b\ll1$ [see Figures~\ref{fig:spectral1} (e) and (f)]. However, %
for $b\gg1$ the best fittings yield to $\alpha = 2.25 - a_{2}/b$ with %
$a_{2} =0.48\pm0.02 $ and $\alpha = 2.7 - a_{4}/b$ with $a_{4} = 1.05\pm0.04$, as $b\rightarrow \infty$ %
for $\beta = 2$ and 4, respectively. It is important to mention that these %
values of $\alpha$ should be regarded as fitting parameters that describe %
the spectral behavior given by equation~(\ref{eq:Pcs1}). Certainly, further %
analytical studies are necessary to see its relation to physical quantities %
like the correlation length exponent~\cite{Evangelou1995,ohtsuki1995}.

In the panels (g)-(i) of Figure~\ref{fig:spectral1} we show the distribution %
$P_{\mathrm{c}}(s)$ of equation~(\ref{eq:pcs2}), for $s\ll 1$. There, we %
present the results for three representative data sets with $b=0.04,\;0.1$, and %
$4$; from left to right. The dashed lines correspond to the Poisson %
distribution,~equation~(\ref{eq:poissonlaw}), while the dotted ones are the %
Wigner--Dyson surmises, equation~(\ref{eq:wignerdyson}). The straight line %
segments are fittings of the numerical data with equation~(\ref{eq:pcs2}). %
In Ref.~\cite{Cuevas2004} the validity of equation~(\ref{eq:pcs2}), for $\beta=1$, %
was verified. In the panel (g) we confirm that %
result. Additionally, here we verify and extend the %
validity of equation~(\ref{eq:pcs2}) to the cases %
$\beta=2$ [panel (h)] and $\beta=4$ [panel (i)], where a good correspondence %
between the numerics and equation~(\ref{eq:pcs2}) is obtained.%

The comparison between our model, equation~(\ref{eq:ModelB}), and the %
analytical estimations, equations~(\ref{eq:P1})-(\ref{eq:P4}), is shown in %
panels (j)-(l) of Figure~\ref{fig:spectral1}. For these figures the parameter %
$a$ is determined by a fitting of the numerical data to the perturbative %
solutions of equations~(\ref{eq:perP1})-(\ref{eq:perP4}), in the regime %
$s \ll 1/a$. The error shown in these panels is the reduced rms which results %
from the fitting.%

In our model, equation~(\ref{eq:ModelB}), the condition $\exp(-\pi^{2}/a)\ll1$ %
means that $b\geq 0.2$ for the range of $b$-values analyzed here. %
Additionally, as shown in panels (g)-(i) of Figure~\ref{fig:spectral1}, if %
$b\geq 2$ the spectral statistics is well described by one of the Wigner--Dyson %
surmises which corresponds to the limit $a\rightarrow0$ in the Nishigaki %
model. Therefore, we chose $b = 0.2$, 0.4 and 1 for the comparison between %
both models. In panels~(j)-(l) of Figure~\ref{fig:spectral1} the dots are the %
numerical data from equation~(\ref{eq:ModelB}), the dotted lines are the %
Wigner--Dyson surmises, the dashed lines are the Poisson distribution, %
equation~(\ref{eq:poissonlaw}), and %
the solid lines correspond to the expressions given in equations~(\ref{eq:P1})%
-(\ref{eq:P4}). In the insets we show the curves of the main panels in semi-log %
scale. In panel (j) the $\beta=1$ case is displayed. For $b=0.2$, 0.4, and 1 %
we found $a= 5.87\pm0.01$, $3.43\pm0.05$, and $1.50 \pm 0.06$, respectively. %
In all these cases the %
agreement between both models is very good. In panel (k) the $\beta=2$ case is %
shown. There, we found for $b=0.2$, 0.4, and 1 values of %
$a = 3.88 \pm 0.02$, $2.01 \pm 0.05$, and $0.92 \pm 0.06$, respectively. %
Again, we observe that both models are in good agreement. %
However, as it is shown in panel (l), the correspondence between both models %
in the symplectic case is guaranteed only for $s \ll 1$; there  for $b=0.2$, %
0.4, and 1 we got $a=4.38 \pm 0.05$, $2.12 \pm 0.02$, and $0.747 \pm 0.032$, respectively.%


\begin{figure*}
\begin{center}
\includegraphics[width=0.70\textwidth]{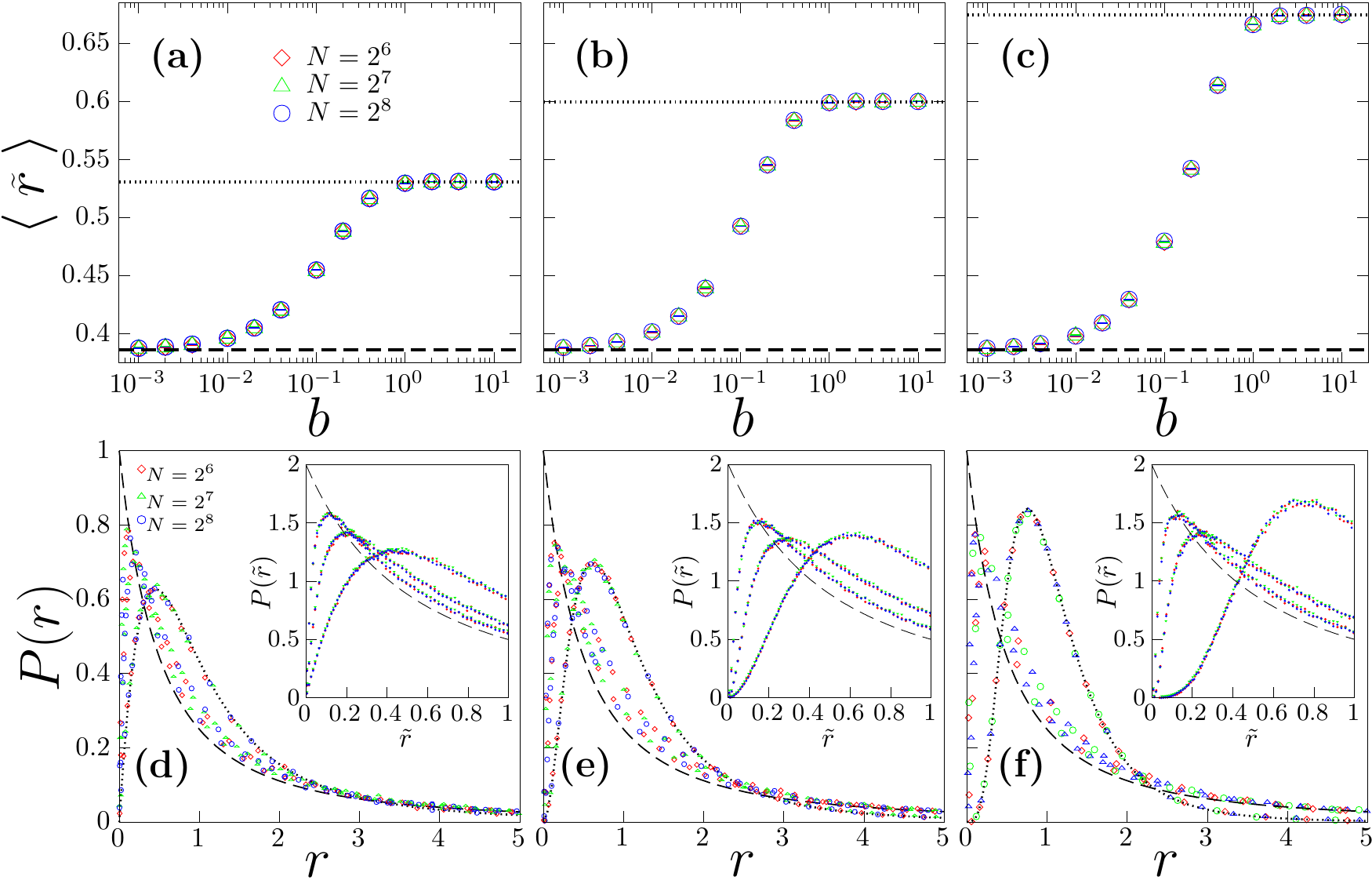}
\caption{%
Spectral statistics of the PBRM model for the ratio of %
consecutive level spacings for $\beta=1$ in left column, panels (a) and (d); %
for $\beta=2$ in middle column, panels (b) and (e); and for $\beta =4$ in %
right column, panels (c) and (f). %
In all panels the dashed lines correspond to the Poisson case and the dotted %
ones correspond to the Wigner--Dyson case, equation~(\ref{eq:PWr}). 
In the panels (d)-(f) the values of %
the parameter $b$ are: $b=0.04$, 0.1 and $4$ (from left to right).
}%
\label{fig:spectral2}
\end{center}
\end{figure*}


Now, we would like to apply the ideas presented in %
subsection~\ref{subsec:ratio} to our PBRM model. First, we calculate the %
average $\langle \tilde{r} \rangle$, equation~(\ref{eq:eratioconsec}), as a %
function of $b$. The results are displayed in Figure~\ref{fig:spectral2}, %
panels (a)-(c), for $\beta=1$, 2, and 4, respectively. In those panels, the %
horizontal dashed lines correspond to the theoretical prediction of $\langle %
\tilde{r} \rangle$ for the Poisson case, while the horizontal dotted lines %
correspond to the respective theoretical values for the corresponding Wigner-%
Dyson ensembles~\cite{Atas2013}. We observe a transition from the Poisson case %
to the Wigner--Dyson cases as a function of $b$, as expected. Also, as we can %
see from this figure, the onset of delocalization occurs at $b\approx0.002$, %
while the Wigner--Dyson regime is reached for $b\approx1$, in all symmetry %
cases.%

In Figure~\ref{fig:spectral2}, panels (d)-(f), the PDF of $r$ [equation~(%
\ref{eq:PWr})] and that of $\tilde{r}$ (in the insets) are shown. The dashed %
lines are the integrable cases [see equations~(\ref{eq:prpoisson}) and (%
\ref{eq:ptrpoisson})], while the dotted ones correspond to the Wigner--Dyson %
expression of equation~(\ref{eq:PWr}). In all these panels, we can see the %
transition from the insulator to metallic phase, that is, when $b=0.04$ the %
points are closer to the dashed line while if $b=4$ the dots are over the dotted %
line, additionally we are including an intermediate case with $b=0.1$. It is %
clear that in the limit $r\rightarrow0$, $P(r)\sim Cr^{\beta}$. It is also %
clear that $P(r)$ as well as $P(\tilde{r})$ allow us to analyze the Poisson to %
Wigner--Dyson transition in the same way as can be done by using $P(s)$, as %
expected. However, the former have the advantage of being easier to compute.%


\section{Conclusions}%
\label{sec:conclusions}%

We have introduced a power-law banded random matrix model for the third of the %
three classical Wigner--Dyson symmetry classes. A detailed analysis of its %
eigenvectors and eigenenergies was presented. This model describes time %
reversal symmetric systems in the presence of strong spin-orbit interactions %
and shows key features of the driven Anderson metal-insulator transition. We %
showed that existing heuristic relations used to describe the multifractal %
properties of the PBRM model in the presence and absence of time reversal %
symmetry, i.e., the ones with $\beta=1$ and 2, respectively, also accounts for %
the PBRM model with $\beta=4$, for some ranges of the model parameters. From %
the statistical analysis, we showed that our results are %
in complete agreement with the corresponding analytical expressions and we %
verified that the metal-insulator transition for the symplectic case also occurs at $\mu=1$. %
In order to present a complete picture of the PBRM model for the three %
classical Wigner--Dyson symmetry classes, we also reproduced previously %
reported results, for $\beta=1$ and $\beta=2$, and also derived some other %
results that have not been shown so far for those symmetry classes.%


\section*{Acknowledgments}

M.C.-N. acknowledges the support to the project Benem\'erita Universidad
Aut\'onoma de Puebla (BUAP) PRODEP 511-6/2019-4354. A.M.M.-A. acknowledges a postdoctoral fellowship from DGAPA-UNAM. J.A.M.-B. thanks support from 
FAPESP (Grant No.~2019/06931-2), Brazil, and
PRODEP-SEP (Grant No.~511-6/2019.-11821), Mexico.


\setcounter{figure}{0}
\begin{appendix}
\section{Multifractal analysis and finite-size scaling analysis}
\label{sec:appendixA}

In this appendix we perform a multifractal analysis (MFA) and a finite-size scaling (FSS) analysis to verify that the PBRM model~(\ref{eq:ModelB}) has a critical point at $\mu=1$. The main idea of the MFA-FSS methods is that the multifractal properties of eigenvectors of a given system do not show FSS dependence when the system is at criticality, and away from criticality the eigenvectors show strong FSS dependence. Based on this idea, several methods have been proposed, see for instance Refs.~\cite{romer2010,romer2011,Mirlin1996,Mirlin2000} and the references 
therein.

Two of the most common quantities that capture the multifractal properties of the eigenvectors of a given system, at criticality, are the so-called participation number $I_{2}$, defined as

\begin{equation}
I_{2}=\left(\sum_{i=1}^{N} |\Psi_{i}|^{4} \right)^{-1},
\label{eq:appendix1}
\end{equation}
with $N$ the system size [c.f~equation.~(\ref{eq:invpn})], and its relative fluctuations

\begin{equation}
\eta = 
\dfrac{\sqrt{\langle(I_{2})^{2}\rangle-\langle I_{2}\rangle}}{\langle I_{2}\rangle}.
\label{eq:appendix2}
\end{equation}
These quantities have been successfully applied to find the critical point in random matrix models, see e.g.~Ref.~\cite{vegaoliveros2019}, and in what follows we will use them together with a MFA-FSS analysis to find the critical 
value of $\mu$ for the PBRM model given by equation~(\ref{eq:ModelB}).

To calculate the equations~(\ref{eq:appendix1}) and~(\ref{eq:appendix2}) we use a direct diagonalization of matrices of size $N=2^{n}$ with $9\leq n\leq 13$, and construct $2^{18-n}$ realizations of the random Hamiltonian~(\ref{eq:ModelB}). For each realization, we calculate $I_{2}$ for an eigenvector window of size $2^{n-3}$ with eigenvalues around the band center $E=0$. Without loss of generality, we fix $b=1$.

\begin{figure}
\centering
\includegraphics[width=0.70\textwidth]{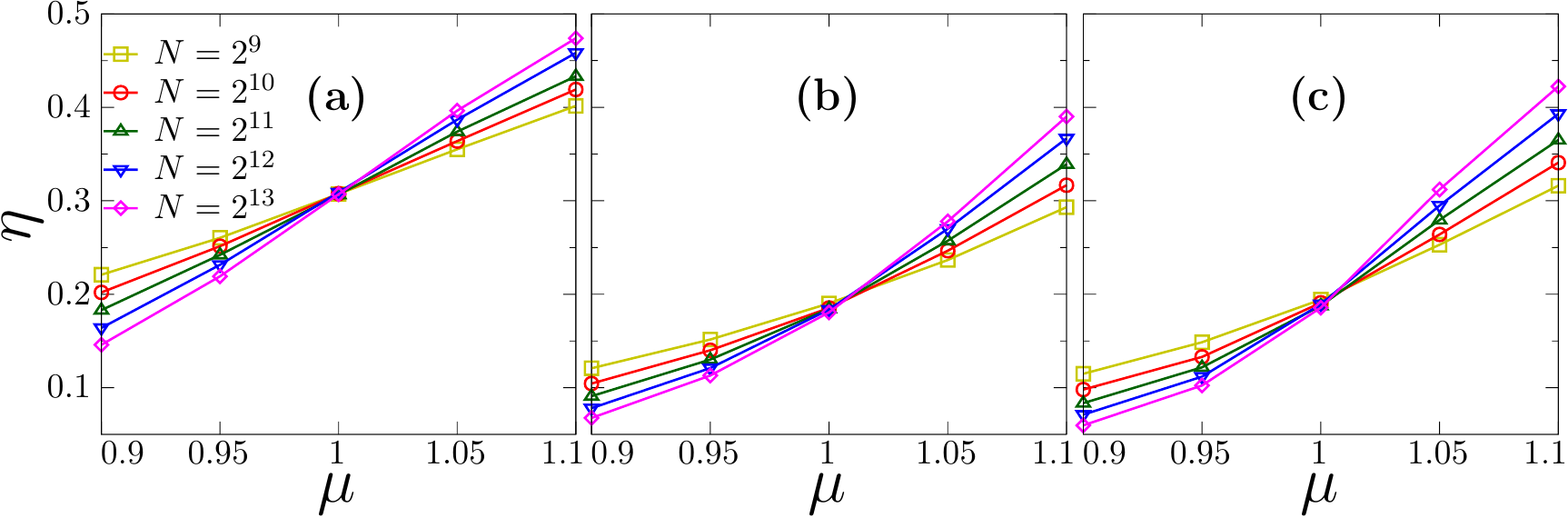}
\caption{Relative fluctuations of the participation number, as function of $\mu$ for the PBRM model of equation~(\ref{eq:ModelB}) with $\beta=1$ (a), $\beta=2$ (b), and $\beta=4$ (c), for several system sizes $N$.}
\label{fig:etavsmu}
\end{figure}

In Figure~\ref{fig:etavsmu}, we show $\eta$
as function $\mu$, for several system sizes $N$, for the three symmetry classes. The $\beta=1$ case is shown in Figure~\ref{fig:etavsmu} (a). It can be observed that for $\mu<1$ and $\mu>1$ the relative fluctuations $\eta$, equation~(\ref{eq:appendix2}), show strong FSS behavior; while for $\mu=1$ all curves have the same value $\eta\approx0.3$, indicating that $\eta(\mu=1)$ is independent of the system size, and therefore the system is at criticality according to the MFA-FSS methods. The $\beta=2$ case is shown in panel (b) of the same figure. There, we use the same MFA-FSS arguments as in the previous case to conclude that $\mu=1$ is a critical point for the model~(\ref{eq:ModelB}) with $\beta=2$. The symplectic case, $\beta=4$, is shown in panel (c). It can be observed that the critical point is also at $\mu=1$, where $\eta(\mu)$ does not show FSS behavior, thus indicating that the PBRM model~(\ref{eq:ModelB}) with $\beta=4$ has a critical point at $\mu=1$.
\begin{figure}[t]
\centering
\includegraphics[width=0.70\textwidth]{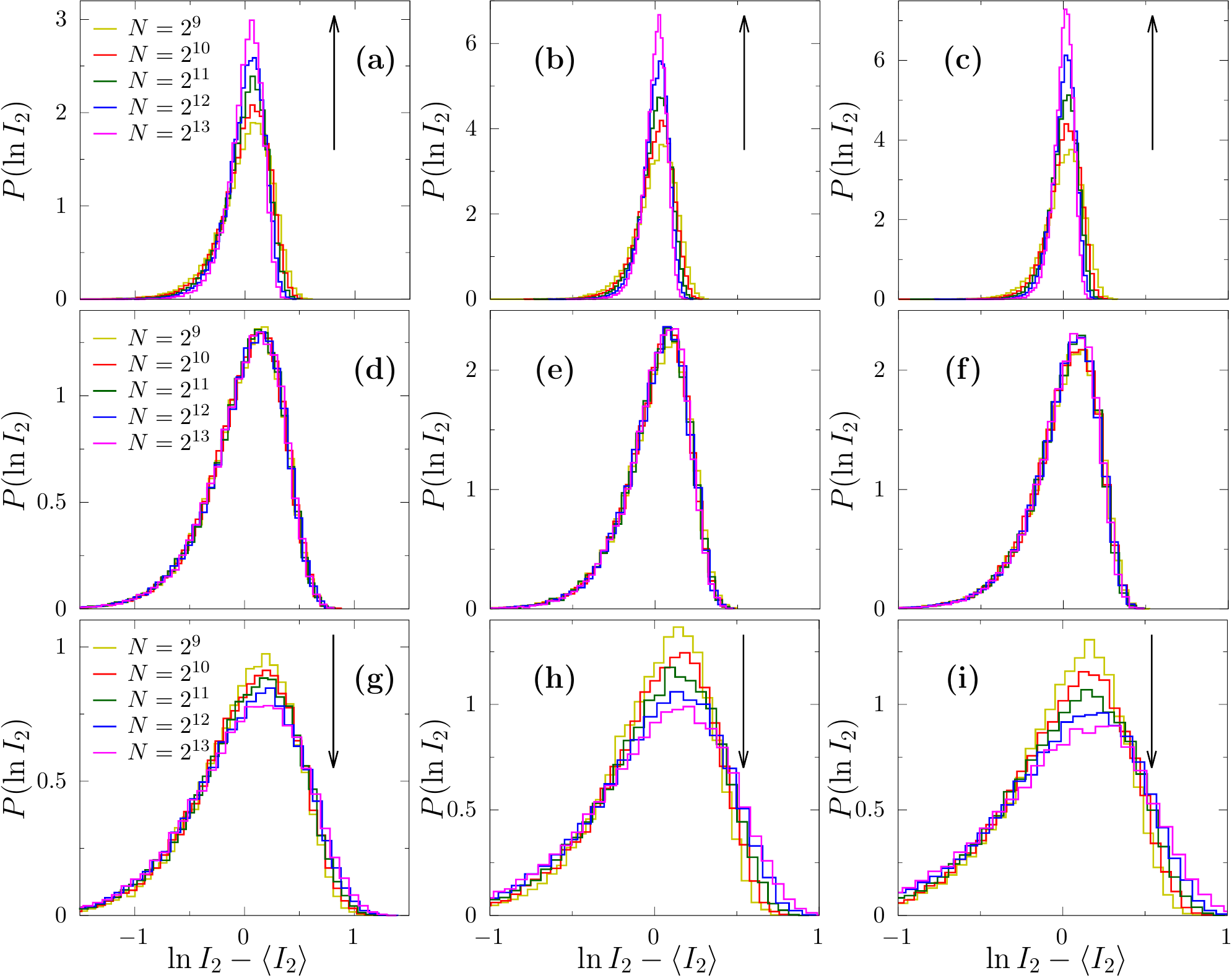}
\caption{Probability density function of the logarithm of the participation number for the PBRM model of equation~(\ref{eq:ModelB}) for several system sizes, for $\beta=1$ [panels (a), (d) and (g)], 
$\beta=2$ [panels (b), (e) and (h)], and, $\beta=4$ [panels (c), (f) and (i)]. In the top row $\mu=0.9$, in middle row $\mu=1$, and in the bottom row $\mu=1.1$. The arrows indicate the system size increase.}
\label{fig:PlnI2}
\end{figure}
For the sake of completeness, in Figure~\ref{fig:PlnI2} we show the probability density function (PDF) of the logarithm of the participation number of equation~(\ref{eq:appendix1}) for several system sizes. The left column [panels (a), (d) and (g)] correspond to $\beta=1$, the middle column [panels (b), (e) and (h)] correspond to $\beta=2$ and the right column [panels (c), (f) and (i)] correspond to $\beta=4$. FSS behavior can be observed in the top raw where $\mu=0.9$. No FSS behavior is shown in the middle row where $\mu=1$. Again FSS behavior is shown at the bottom row where $\mu=1.1$. Those behaviors indicate that the PBRM model of equation~(\ref{eq:ModelB}) has a critical point at $\mu=1$ for the three symmetries classes studied here, namely, the orthogonal ($\beta=1$), the unitary ($\beta=2$) and the symplectic ($\beta=4$).

\end{appendix}


\bibliography{biblio}
\bibliographystyle{elsarticle-num}


\end{document}